\documentclass[12pt]{iopart}
\usepackage{graphicx}
\usepackage{cite}
\def\U#1{{%
\def\O{\mbox{O}}
\def\u{\mbox{u}}
\mathcode`\u=\mu
\mathcode`\O=\Omega
\mathrm{#1}}}
\def\sub#1{_{\scriptsize\mbox{#1}}}
\begin{document}

\title[]{Low-power threshold gas discharge by enhanced local electric field in electromagnetically-induced-transparencylike metamolecules}

\author{Yasuhiro Tamayama$^*$ and Ryosuke Yamada}

\address{Department of Electrical, Electronics and
Information Engineering, Nagaoka University of
Technology, 1603-1 Kamitomioka, Nagaoka, Niigata 940-2188, Japan}
\ead{tamayama@vos.nagaokaut.ac.jp}
\vspace{10pt}
\begin{indented}
\item[]\today
\end{indented}

\begin{abstract}
To realize efficient nonlinear metamaterials, we investigate a method for enhancing the local electric field in a metamolecule composed of two radiatively coupled cut-wire resonators where resonance of the cut-wire resonators and low-group-velocity propagation of an incident electromagnetic wave simultaneously occur. Numerical analysis shows that the local electric field in the metamolecule can be enhanced by decreasing the electrode size and the gap of the capacitor structure of the cut-wire resonators while keeping the equivalent electrical circuit parameters of the metamolecule constant. We measure and compare the threshold incident power for a gas discharge in the metamolecule fabricated in our previous study and that in the metamolecule with shrunken capacitor structures. The experiment reveals that 
shrinking the capacitor structure while keeping the resonance frequency of the metamolecule
decreases the threshold incident power for a gas discharge and may increase the gas pressure where the threshold incident power is minimum. Further development of this work will enable us to realize efficient nonlinear metamaterials that have atmospheric-pressure gas as a nonlinear element.
\end{abstract}

%
%
%
%
%

\section{Introduction}

Extensive investigation has been done on the nonlinear response of metamaterials because they can be used to generate nonlinear phenomena efficiently~\cite{lapine_14_rmp,krasnok_18_mat_tod}. When an electromagnetic wave is incident on a resonant metamaterial, the electric/magnetic energy is compressed into a small volume in the metamaterial. Nonlinear phenomena can then be efficiently generated by placing a nonlinear element in this small volume. To realize nonlinear metamaterials, pn junctions (or Schottky junctions)~\cite{rose_11_prl,kanazawa_11_apl,nakanishi_12_apl,filonov_16_apl,tamayama_13_prb,kozyrev_14_apl,agaoglou_14_bif,}, 
superconductors~\cite{zhang_d_16_prb,savinov_16_apl}, vanadium dioxide~\cite{liu_12_nat}, gallium arsenide~\cite{fan_13_prl,keiser_17_apl}, plasma~\cite{iwai_15_pre,gregorio_16_psst,fantini_19_jap}, and the nonlinear magnetic component of the Lorentz force in metals~\cite{wen_17_prl} are used in the microwave and terahertz regions, and the nonlinearity of metals and dielectrics
is used in the optical region~\cite{czaplicki_13_prl,celebrano_15_nat_nano,obrien_15_nat_mat,yang_15_nl,liu_16_nl,shorokhov_16_nl,alberti_16_apb,gennaro_16_nl,metzger_16_apb,su_16_sci_rep,wang_17_acsphoton,wolf_17_sci_rep,yang_17_acsphoton,ren_12_nat_com,guddala_16_ol,lawrence_18_nl,noskov_12_prl}.

Resonance in a metamaterial enhances the local electric field in it; thus, the electric field enhancement is suppressed if the nonlinear elements have losses. To prevent this suppression, a nonlinear dielectric resonator with low loss~\cite{yang_15_nl,liu_16_nl,shorokhov_16_nl} and the magnetic component of the Lorentz force~\cite{wen_17_prl} should be used as nonlinear elements, for example. A gas discharge can also be used as a low-loss nonlinear element, as demonstrated in our previous study~\cite{tamayama_18_apl}. We believe that a gas discharge can be a useful nonlinear element especially in a low-frequency region. When a gas discharge is used as a nonlinear element, strong enhancement of the local electric field is essential for generating nonlinear phenomena because these phenomena never occur below the threshold local electric field for a gas discharge. Thus far, we have demonstrated that a local electric field can be strongly enhanced using a metamaterial composed of two radiatively coupled resonators where the resonance of the resonators occurs simultaneously with the low-group-velocity propagation of an incident electromagnetic wave~\cite{tamayama_15_prb,tamayama_17_jap}.
Although the effects of the resonance and the low-group-velocity propagation on the local electric field enhancement are not clearly separable in resonant metamaterials and the group velocity cannot be defined unambiguously in single-layer metamaterials (metasurfaces), we have confirmed that the local electric field in resonant metamaterials can be strongly enhanced by causing large group delay.
In those studies, only equivalent electrical circuit parameters of the metamaterial were taken into account to enhance the local electric field. To further enhance the local electric field for efficient generation of nonlinear phenomena, this study investigates the dependence of the local electric field enhancement on the capacitor structure in a metamaterial where the resonance and the low-group-velocity propagation simultaneously occur. We also evaluate the threshold incident power for a gas discharge in the designed metamaterial to clarify the influence of the capacitor structure on the threshold incident power.

\section{Design of metamolecular structure for local electric field enhancement}

First, we consider the relationship between the local electric field and the capacitor structure of the resonators to further enhance the local electric field in a metamaterial where the resonance and the low-group-velocity propagation simultaneously occur. Assuming that the capacitor structure can be regarded as a parallel plate capacitor for simplicity, $q=\varepsilon_0 SE$ is satisfied, where $q$ is the stored charge, $\varepsilon_0$ is the permittivity in vacuum, $S$ is the plate area of the capacitor, and $E$ is the electric field in the capacitor. This equation shows that $E$ can be enhanced by reducing $S$ while keeping $q$ constant. However, if only $S$ is reduced, $q$ varies because of the resonance frequency shift caused by the variation of the capacitance $C = \varepsilon_ 0 S/d$ ($d$ is the separation of the parallel plates). Both $S$ and $d$ should be reduced to keep $q$ constant and to enhance $E$. It follows that the local electric field in the metamaterial can be enhanced by reducing the electrode size and spacing of the capacitor structures while keeping the equivalent electrical circuit parameters of the metamaterial unchanged. Although it is well known that a strong electric field can be generated near a sharp electrode, $S/d$ should be kept constant to simultaneously generate the resonance and the low-group-velocity propagation at a fixed frequency in this case.

\begin{figure}[tb]
\begin{center}
\includegraphics[scale=0.9]{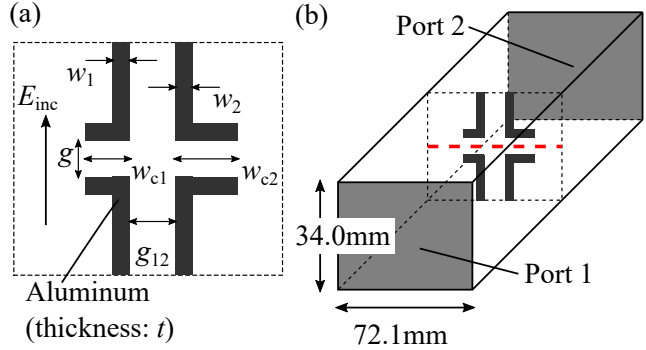}
\caption{Schematic of (a) the metamolecular structure and (b) the simulation system. The incident electric field is along the vertical direction. Ports 1 and 2 are shaded. A perfect electric conductor boundary condition is applied to the other boundaries.}
\label{fig:structure}
\end{center}
\end{figure}

We verify the above discussion through numerical simulations. Figure \ref{fig:structure}(a) shows the structure of a metamolecule used in this study. This structure is composed of two radiatively coupled cut-wire resonators with slightly different resonance frequencies. We showed in our previous study~\cite{tamayama_15_prb} that the electric field in the gaps of the resonators is strongly enhanced by simultaneous generation of resonance of the cut-wire resonators and low-group-velocity propagation of an incident electromagnetic wave at a frequency that is about the average of the resonance frequencies of the cut-wire resonators. The geometrical parameters of the metamolecule in that study (which is called metamolecule A hereafter) were $w_1 = w_2 = 3.0\,\U{mm}$, $w\sub{c1} = 5.7\,\U{mm}$, $w\sub{c2}=6.0\,\U{mm}$, $g=1.0\,\U{mm}$, $g_{12}= 3.0\,\U{mm}$, and $t= 1.0\,\U{mm}$. Here, we apply the above theory to this metamolecular structure. The gap and the electrode size of the capacitor structure should be reduced to enhance the local electric field; thus, we set $g=0.3\,\U{mm}$ and $w\sub{c1}=1.0\,\U{mm}$, which are determined by a restriction in our fabrication process described below. Assuming that the gap structure can be regarded as a parallel plate capacitor, $w\sub{c1}$ should be equal to $5.7\,\U{mm} / (1.0\,\U{mm} / 0.3\,\U{mm}) = 1.7\,\U{mm}$ to keep $C$ constant. However, we set $w\sub{c1}=1.0\,\U{mm}$ to further reduce $S$ and to further enhance $E$. Instead, we vary the inductances of the cut-wire resonators, which depend on $w_1$ and $w_2$, so that the resonance frequencies are almost equal to those in metamolecule A. We then compare the electric field enhancement factors in metamolecule A and the designed metamolecule on the basis of the present theory.

We numerically analyzed the electromagnetic responses of these two metamolecules using COMSOL Multiphysics to calculate their electric field enhancement factors. The geometry of the simulation system is shown in Fig.\,\ref{fig:structure}(b). The conductivity of aluminum was assumed to be $3.0\times 10^5\,\U{S/m}$ by reference to our previous study~\cite{tamayama_17_jap}. A perfect electric conductor boundary condition was applied to the boundaries shown in white because the metamolecule was placed in a rectangular waveguide in the experiment described below. An incident electromagnetic wave with mode TE${}_{10}$ was radiated from Port 1, and the transmitted wave was detected at Port 2. 
(Note that the walls of the rectangular waveguide behave almost as periodic boundaries in this condition. 
It follows that the electromagnetic response of the metamolecule in this case is similar to that of a two-dimensional periodic array of this metamolecule to a plane electromagnetic wave.
In fact, we have numerically confirmed that a similar phenomenon is observed in metasurfaces composed of a periodic array of this type of metamolecule~\cite{tamayama_14_prb}.)
A preliminary simulation showed that the resonance frequency of a cut-wire resonator with $g=0.3\,\U{mm}$ and $w\sub{c1}=1.0\,\U{mm}$ becomes about 3\,GHz when $w_1 = 1.0\,\U{mm}$. We set $w_2$ to 1.0\,mm to make the structures of the two cut-wire resonators similar and to make the radiative coupling between the two resonators strong. The condition for maximizing the local electric field enhancement was examined by varying $w\sub{c2}$ according to the theory described in our previous paper~\cite{tamayama_14_prb}. We determined $g_{12}$ so that the transmission peak frequency was about the average of the resonance frequencies of the two cut-wire resonators.

\begin{figure}[tb]
\begin{center}
\includegraphics[scale=0.68]{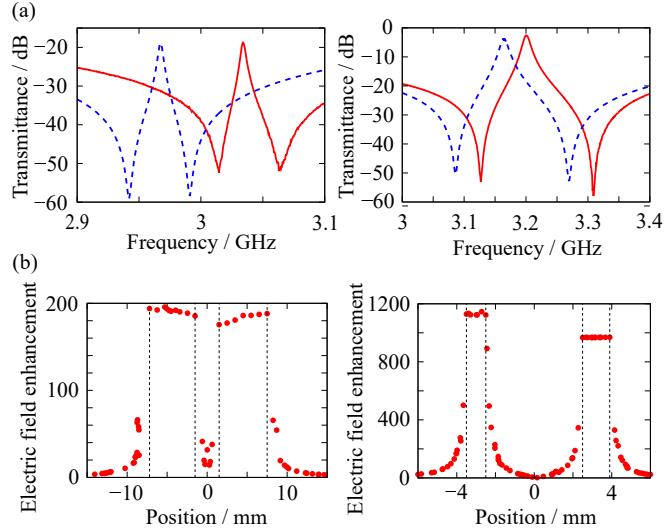}
\caption{(a) Transmission spectra (solid curve: experiment; dashed curve: simulation) and (b) distributions of the calculated electric field enhancement factor along the red bold dashed line in Fig.\,\ref{fig:structure}(b) at the transmission peak frequency for (left) metamolecule A and (right) metamolecule B. Vertical dashed lines in (b) represent the positions of the capacitor structures.}
\label{fig:linear}
\end{center}
\end{figure}

Figure \ref{fig:linear}(a) shows the numerically calculated transmission spectra of metamolecule A and the designed metamolecule B with $w\sub{c2}=1.4\,\U{mm}$ and $g_{12} = 5.0\,\U{mm}$, which is the condition for maximizing the local electric field. An electromagnetically-induced-transparencylike (EIT-like) transmission peak is observed at 2.967\,GHz and 3.164\,GHz for metamolecules A and B, respectively. Strong enhancement of the local electric field occurs at the transmission peak frequency. Figure \ref{fig:linear}(b) shows the calculated electric field distributions in the metamolecules along the red bold dashed line in Fig.\,\ref{fig:structure}(b) at the transmission peak frequency. The maximum electric field enhancement factors in metamolecules A and B are 196 and 1146, respectively. The ratio of these values is 5.8, which roughly agrees with the ratio of the electrode sizes, $5.7\,\U{mm} / 1.0\,\U{mm} = 5.7$. Although the above theory for enhancing the local electric field based on the characteristic of a parallel plate capacitor is very simple, the theory roughly agrees with the result of numerical analysis. The result confirms that the local electric field can be further enhanced by reducing the electrode size and the gap of the capacitor structure in EIT-like metamolecules while keeping the effects of the resonance and the low-group-velocity propagation. 

Here, we discuss the relationship between the group delay and the electric field enhancement factor in the metamolecules. The numerically calculated group delay in metamolecule A (metamolecule B) at the transmission peak frequency was 78\,ns (22\,ns), and the ratio of the electric field enhancement factor in metamolecule A (metamolecule B) to that in a meta-atom composed of either of the cut-wire resonators in metamolecule A (metamolecule B) was 6.1 (6.5). These results show that the electric field enhancement factor in the metamolecules and the ratio of the electric field enhancement factor in the metamolecule to that in the meta-atom composed of either of the cut-wire resonators do not depend solely on the group delay. Although the local electric field enhancement factor in resonant metamaterials can be increased by introducing low-group-velocity propagation~\cite{tamayama_15_prb}, the electric field enhancement factor does not depend only on the group delay but also on the capacitor structure.

\section{Measurement of threshold incident power for gas discharge}

Next, we investigate how enhancing the local electric field by shrinking the capacitor structure influences the threshold incident power for a gas discharge in the metamolecules. 
The designed metamolecule was made from an aluminum plate with a thickness of 1.0\,mm using wire electrical discharge machining. The fabricated metamolecule was placed in a rectangular waveguide with cross-sectional dimensions of $72.1\,\U{mm} \times 34.0\,\U{mm}$. Before evaluating the threshold incident power for a gas discharge, we measured the transmission characteristics of the metamolecule to find the transmission peak frequency that maximized the local electric field enhancement. The transmission spectra of metamolecules A and B measured using a network analyzer are shown in Fig.\,\ref{fig:linear}(a). An EIT-like transmission peak is observed at 3.034\,GHz (3.200\,GHz) for metamolecule A (metamolecule B). The group delay at the transmission peak frequency was measured to be 81\,ns (26\,ns) for metamolecule A (metamolecule B). These results roughly agree with the results of the numerical analysis. Then, we measured the threshold incident power for gas discharges in the metamolecules. The rectangular waveguide, in which the metamolecule was placed, was put in an acrylic vacuum chamber that was filled with Ar gas with pressure $p\sub{Ar}$. An incident electromagnetic wave was generated by a signal generator. The generated signal was amplified, fed into the waveguide via a coaxial-waveguide transformer, and incident on the metamolecule. The transmitted wave was detected by an ultra-wideband dipole antenna~\cite{lule_05_mop}, which enabled us to measure the transmittance of the metamolecule and to observe the gas discharge in the metamolecule simultaneously~\cite{tamayama_18_apl}. The detected signal was fed into a spectrum analyzer via an attenuator to measure the power of the transmitted wave. In the measurement of the transmitted power, the incident power was swept from low to high and then from high to low. The transmittance was calculated as the ratio of the power before the attenuator to that after the amplifier.

\begin{figure}[tb]
\begin{center}
\includegraphics[scale=0.69]{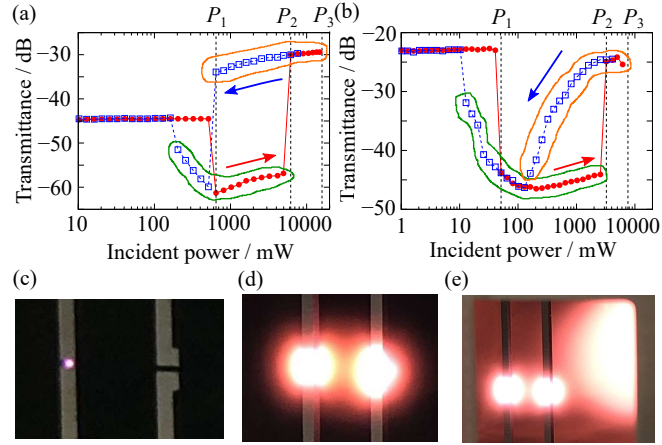}
\caption{Incident power dependence of the transmittance of (a) metamolecule A and (b) metamolecule B. Red circles (blue squares) represent the measured values when the incident power is swept from low to high (high to low). The values of $P_1$, $P_2$, and $P_3$ for metamolecule A (metamolecule B) are 0.65\,W, 6.2\,W, and 16\,W (0.051\,W, 3.2\,W, and 7.6\,W), respectively. Lines that connect the measured data are guides to the eye. A gas discharge appears in the gap of the cut-wire resonator with higher resonance frequency in the region surrounded by the green line and in both gaps of the cut-wire resonators in the region surrounded by the orange line. Photographs of metamolecule B in the regions surrounded by the (c) green line and (d) orange line. Photograph for (e) an incident power higher than $P_3$. The electromagnetic wave is incident from the back on the metamolecule in the photograph.}
\label{fig:trans_power}
\end{center}
\end{figure}

Figures \ref{fig:trans_power}(a) and \ref{fig:trans_power}(b) show the incident power dependence of the transmittance of metamolecules A and B for $p\sub{Ar} =1\,\U{kPa}$ at their transmission peak frequencies, 3.029\,GHz and 3.195\,GHz, respectively. The transmission peak frequencies in this experiment are slightly different (within several MHz) from the experimental result in Fig.\,\ref{fig:linear}(a), which is caused by the slight difference in the positions of the metamolecules in the waveguide. For both metamolecules, the transmittance exhibits a constant value for low incident power and sharply decreases when the incident power exceeds $P_1$. This observation shows that the gas discharge occurs in the gap of the cut-wire resonator with higher resonance frequency, as shown in Fig.\,\ref{fig:trans_power}(c). The value of $P_1$ reflects the electric field enhancement caused by the resonance, the low-group-velocity propagation, and the shrinking of the capacitor structure. The value of $P_1$ for metamolecule A is $13$ $(= 3.6^2 )$ times higher than for metamolecule B. This confirms that the threshold incident power for the gas discharge is decreased by shrinking the capacitor structure, while the ratio of the threshold incident power is smaller than the value estimated from the numerically calculated electric field enhancement factor, $5.8^2$. This discrepancy will be discussed with the pressure dependence of the threshold incident power below. 

Although measuring $P_1$ is sufficient for evaluating the local electric field enhancement owing to the resonance, the low-group-velocity-propagation, and the shrinking of the capacitor structure, we also measured the incident power dependence of the transmittance at higher power to further investigate the gas discharge in the metamolecules. When the incident power exceeds $P_2$, the transmittance rapidly increases. This is because gas discharges occur in both gaps of the cut-wire resonators simultaneously, as shown in Fig.\,\ref{fig:trans_power}(d). Both the resonance transmission dips disappear because of the gas discharge; thus, the transmittance increases. The value of $P_2$ should reflect the electric field enhancement caused by the resonance of the cut-wire resonator meta-atoms. The ratio of $P_2$ to $P_1$ for metamolecule A [metamolecule B] is $9.5$ $(= 3.1^2 )$ [$63$ $(= 7.9^2 )$], and the ratio of the numerically calculated electric field enhancement factor in metamolecule A (metamolecule B) to that in a meta-atom composed of either of the cut-wire resonators in metamolecule A (metamolecule B) is 6.1 (6.5). The value of $\sqrt{P_2 / P_1}$ is not significantly different from the ratio of the electric field enhancement factors for metamolecule B, while $\sqrt{P_2 / P_1}$ is only half the electric field enhancement factor ratio for metamolecule A. The value of $P_2$ may decrease under the influence of the gas discharge~\cite{liu_ch_14_jap} in the gap of the cut-wire resonator with a higher resonance frequency in metamolecule A because the distance between its two cut-wire resonators is relatively close compared with that in metamolecule B. When the incident power exceeds $P_3$, a gas discharge occurs in a space between the metamolecule and the excitation source in the waveguide, as shown in Fig.\,\ref{fig:trans_power}(e). The transmitted power drifted for a while in this state in spite of a constant incident power; thus, the transmittance in this state is not shown in Figs.\,\ref{fig:trans_power}(a) and \ref{fig:trans_power}(b). 
This phenomenon is not associated with the electric field enhancement in the EIT-like metamolecule, so we do not investigate it further.
The gas discharge in the space in front of the metamolecule disappears when the incident power drops below a certain value. The transmittance is plotted again from this point as blue squares in Figs.\,\ref{fig:trans_power}(a) and \ref{fig:trans_power}(b). The dependence of the transmittance on the incident power when the incident power is swept from high to low is different from that when the incident power is swept from low to high. This is because the incident power necessary to generate the gas discharge is higher than the incident power necessary to maintain the gas discharge, as discussed in our previous paper~\cite{tamayama_17_jap}.

\begin{figure}[tb]
\begin{center}
\includegraphics[scale=0.69]{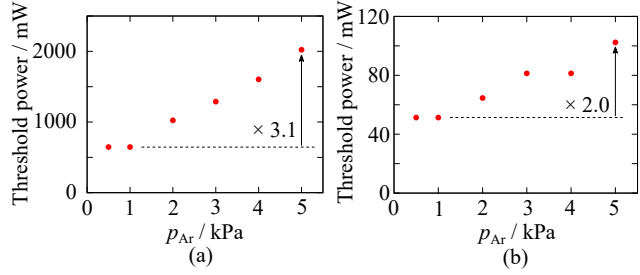}
\caption{$p\sub{Ar}$ dependences of the threshold incident power $P_1$ for the gas discharge in (a) metamolecule A and (b) metamolecule B.}
\label{fig:thre}
\end{center}
\end{figure}

Finally, we measured the $p\sub{Ar}$ dependences of the threshold incident power $P_1$ for the gas discharge to better understand the influence of the capacitor structure on the threshold incident power. The measured results for metamolecules A and B are shown in Fig.\,\ref{fig:thre}. The threshold power increases with $p\sub{Ar}$ in the measured pressure range for both metamolecules, while the ratio of the threshold power for 5\,kPa to that for 0.5\,kPa is different. This ratio is 3.1 for metamolecule A and 2.0 for metamolecule B. We infer from this result that the value of $p\sub{Ar}$ where the threshold incident power is minimum (which we define as $p\sub{min}$) for metamolecule B may be higher than for metamolecule A. If this is true, $p\sub{min}$ can be increased by shrinking the capacitor structure while keeping the resonance frequency, which is consistent with Paschen's law. 
The dependence of the threshold incident power on $p\sub{Ar}$ seems to depend on the gap $g$; thus, it is difficult to quantitatively compare the numerically calculated electric field enhancement factors with the measured threshold incident powers at a certain pressure. However, at a higher pressure than $p\sub{min}$, the threshold incident power definitely decreases when the capacitor structure is shrunk, which is an important observation.
Assuming that Paschen's law can be applied simply, $p\sub{min}$ increases to 100 times the value for metamolecule B when $g$ decreases to the order of micrometers. Metamolecules with overall sizes of tens of millimeters and gaps of micrometers~\cite{Cohick_20_psst} may enable us to realize efficient nonlinear metamaterials that include atmospheric-pressure gas as a nonlinear element.

\section{Conclusion}

We investigated a method for enhancing the local electric field in an EIT-like metamolecule composed of two radiatively coupled cut-wire resonators by shrinking their capacitor structures while keeping the resonance frequencies. We also investigated the influence of shrinking the capacitor structure on the threshold incident power for a gas discharge. The local electric field enhancement caused by the shrinking of the capacitor structure is compatible with that based on the resonance and the low-group-velocity propagation in the EIT-like metamolecule; thus, the local electric field can be further enhanced than in our previous study. Numerical simulation confirmed that the local electric field enhancement factor was increased by shrinking the capacitor structure while keeping the resonance frequency of the metamolecule, and that the ratio of the electric field enhancement factors in metamolecules A and B agreed with the ratio of the electrode sizes of their capacitor structures. This agrees with a parallel plate capacitor model. At a higher pressure than $p\sub{min}$, the threshold incident microwave power for a gas discharge decreased owing to the electric field enhancement caused by the shrinking of the capacitor structure. The $p\sub{Ar}$ dependence of the threshold incident power showed that $p\sub{min}$ may be increased by shrinking the capacitor structure. Further development of this work may enable us to efficiently generate nonlinear responses of metamaterials based on gas discharges at atmospheric pressure. Such nonlinear metamaterials would widen application of nonlinear electromagnetics. Further development of this work would also yield a method for efficiently generating atmospheric-pressure microplasma using microwaves.

\section*{Data availability statement}
The data that support the findings of this study are available upon reasonable request from the authors.

\section*{Acknowledgment}
This research was supported by JSPS KAKENHI Grant Number JP18H03690. 

\section*{References}
\bibliographystyle{unsrt}

\end{document}